\documentstyle[12pt,a4]{article}

\begin{document}

\begin{titlepage}
\vspace*{3cm}
\centerline{\bf Duality of Reggeon Interactions in QCD{$^*$}}
\date{}
\begin{center}
L.\,N.\ Lipatov {$^{\dagger}$\\
Petersburg Nuclear Physics Institute,\\
Gatchina, 188350, St.Petersburg, Russia}
\end{center}
\vskip 15.0pt
\centerline{\bf Abstract}

\vspace{.2cm}
\noindent
The duality symmetry of the hamiltonian and integrals of
motion for reggeon interactions in the multi-colour QCD is
formulated as an integral equation for the wavefunction of
compound states of reggeized gluons, which together with the
property of holomorphic factorization allows one to reduce the
Odderon problem in QCD to the solution of the one-dimensional
Schr\"odinger equation.
\vskip 2cm \hrule
\vskip .4cm \noindent
\noindent $*$ {Talk given at DIS 98, Brussels, April 1998}
\vskip .4cm \noindent
\noindent ${\dagger}$ {\it Supported by CRDF, INTAS and RFFI
grants}
\vfill
\end{titlepage}

In the multi-colour QCD the holomorphic hamiltonian $h$ [1] for compound
states of $n$ reggeized gluons with their momenta $p_{k}=i\partial _{k}$
and
relative coordinates $\rho _{k,l}=\rho _k - \rho _l$

\begin{equation}
h=\sum_{k=1}^{n}\left( 2\,\ln \,p_{k}+p_{k}^{-1}\,(\ln \,\rho
_{k,k+1})\,\,p_{k}+p_{k+1}^{-1}\,(\ln \,\rho _{k,k+1})\,\,p_{k+1}+2\,\gamma
\right)\,\,,\,\,\,\gamma =\psi (1)
\end{equation}
and the integrals of motion [2]

\begin{equation}
q_{r}=\sum_{1 \leq i_{1}<i_{2}<...<i_{r} \leq n}\rho _{i_{1},i_{2}}\,\rho
_{i_{2},i_{3}...}\rho
_{i_{r},i_{1}\,}\,p_{i_{1}}p_{i_{2}}...p_{i_{r}}\,,\,\,p_k=i\partial_k
\end{equation}
are invariant under the cyclic permutation of gluon indices $i\rightarrow
i+1 $, corresponding to the Bose symmetry of the reggeon wavefunction at
$%
N_{c}\rightarrow \infty $. It is remarkable, that these operators are
invariant also under the more general canonical transformation:

\begin{equation}
\rho _{i-1,i}\rightarrow p_{i}\rightarrow \rho _{i,i+1}\,,\,i=1,2,...n
\end{equation}
combined with reversing the order of the operator multiplication.

Note that the supersymmetry corresponds to an analogous generalization of
translations to super-translations. Furthermore, the Kramers-Wannier
duality
in the Ising model and the popular electro-magnetic duality
$\overrightarrow{%
E}\leftrightarrow \overrightarrow{H}$ [3] can be considered as similar
canonical transformations.

The above duality symmetry is realized as an unitary transformation only
for
the vanishing total gluon momentum $\overrightarrow{p}=0$. The 
wavefunction 
$\psi _{m,\widetilde{m}}$ of the colourless state with
$\overrightarrow{p}=0$
can be written in terms of the eigenfunction $f_{m\widetilde{m}}$ of the
commuting set of operators $q_{k}$ and $q_{k}^{*}$ for $k=2,3,...n$ as
follows

\begin{equation}
\psi _{m,\widetilde{m}}(\overrightarrow{\rho _{12}},\overrightarrow{\rho
_{23}},...\overrightarrow{\rho _{n1}})=\int \frac{d^2\rho _0}{2\,\pi
}\,f_{m,%
\widetilde{m}}(\overrightarrow{\rho _1},\overrightarrow{\rho _2},...%
\overrightarrow{\rho _n};\overrightarrow{\rho _0})\,.
\end{equation}

It is the highest weight component of the M\"{o}bius group unitary
representation with the conformal weights $m=1/2+i\nu +n/2$ and $\,%
\widetilde{m} =1/2+i\nu -n/2$ [1].

Taking into account the known hermiticity properties of the hamiltonian $h$
and the fact, that $\psi _{m,\widetilde{m}}$ is an eigenfunction of the
integrals of motion $q_{n}$ and $q_{n}^{*}$ with the eigenvalues
$%
\lambda _{m}$ and $\lambda _{m}^{*}$ [1], one can formulate the duality
symmetry of the reggeon interactions in the form of the integral equation: 
\begin{equation}
\psi _{m,\widetilde{m}}(\overrightarrow{\rho _{12}},...\overrightarrow{\rho
_{n1}})=\left| \lambda _{m}\right| \,2^{n}\,\int \,\prod_{k=1}^{n-1}\frac{%
d^{2}\rho _{k-1,k}^{\prime }}{2\pi }\,\prod_{k=1}^{n}\frac{e^{i%
\overrightarrow{\rho _{k,k+1}}\,\overrightarrow{\rho _{k}^{\prime }}}}{%
\left| \rho _{k,k+1}^{\prime }\right| ^{2}}\,\psi _{\widetilde{m},m}^{*}(%
\overrightarrow{\rho _{12}^{\prime }},...,\overrightarrow{\rho
_{n1}^{\prime
}})\,.
\end{equation}

In the case of the Odderon the M\"{o}bius invariance fixes the solution of
the Schr\"{o}dinger equation

\begin{equation}
f_{m,\widetilde{m}}(\overrightarrow{\rho _1},\overrightarrow{\rho _2},%
\overrightarrow{\rho _3};\overrightarrow{\rho _0})=\left( \frac{\rho
_{12}\,\rho _{23}\,\rho _{31}}{\rho _{10}^2\,\rho _{20}^2\,\rho _{30}^2}%
\right) ^{m/3}\left( \frac{\rho _{12}^{*}\,\rho _{23}^{*}\,\rho
_{31}^{*}}{%
\rho _{10}^{*2}\,\rho _{20}^{*2}\,\rho _{30}^{*2}}\right) ^{\widetilde{m}%
/3}f_{m,\widetilde{m}}(\overrightarrow{x})\,
\end{equation}
up to an arbitrary function $f_{m,\widetilde{m}}(\overrightarrow{x})$ of
one
complex anharmonic ratio $x=\frac{\rho _{12}\,\rho _{30}}{\rho _{10}\,\rho
_{32}}$. Therefore its wavefunction 
$\psi _{m,\widetilde{m}}(\overrightarrow{\rho 
_{12}},\overrightarrow{\rho _{23}},\overrightarrow{\rho _{31}})$ at 
$\overrightarrow{q}=0$ can be written as follows

\begin{equation}
\psi _{m,\widetilde{m}}(\overrightarrow{\rho _{12}},
\overrightarrow{\rho _{23}},\overrightarrow{\rho _{31}})=\left( 
\frac{\rho _{23}%
}{\rho _{12}\rho _{31}}\right) ^{m-1}\left( \frac{\rho _{23}^{*}}{\rho
_{12}^{*}\rho _{31}^{*}}\right) ^{\widetilde{m}-1} \chi
_{m,\widetilde{m}}(%
\overrightarrow{z})\,\,,\,\,\,z=\frac{\rho _{12}}{\rho _{32}}\,,
\end{equation}
where

\begin{equation}
\chi _{m,\widetilde{m}}(\overrightarrow{z})=\int \frac{d^{2}x\,\,f_{m,%
\widetilde{m}}(\overrightarrow{x})}{2\,\pi \left| x-z\right| ^{4}}\,\left( 
\frac{(x-z)^{3}}{x(1-x)}\right) ^{2m/3}\left( \frac{(x^{*}-z^{*})^{3}}{%
x^{*}(1-x^{*})}\right) ^{2\widetilde{m}/3}.
\end{equation}

Choosing a certain phase for the Odderon wavefunction one can present the
duality equation in the pseudo-differential form: 
\begin{equation}
z(1-z)\,\,p^{1+m}\,z^{*}(1-z^{*})\,(p^{*})^{1+\widetilde{m}}\,\varphi _{m,%
\widetilde{m}}(\overrightarrow{z})=\left| \lambda _{m}\right| \,\left(
\varphi _{m,\widetilde{m}}(\overrightarrow{z})\right) ^{*}\,,
\end{equation}
where $p=i\partial ,\,p^{*}=i\partial ^{*}$ and 
\begin{equation}
\varphi _{m,\widetilde{m}}(\overrightarrow{z})\sim \left( z(1-z)\right)
^{m}\left( z^{*}(1-z^{*})\right) ^{\widetilde{m}}\,\chi
_{1-m,1-\widetilde{m}%
}(\overrightarrow{z})\,\,,\,\,\varphi
_{1-m,1-\widetilde{m}}(\overrightarrow{%
z})=\varphi _{m,\widetilde{m}}^{*}(\overrightarrow{z})\,.
\end{equation}

The duality equation in the integral form is written below

\begin{equation}
\varphi _{m,\widetilde{m}}(\overrightarrow{x})=\left| \lambda _{m}\right|
\,%
\frac{i^{\widetilde{m}-m}\Gamma (-\widetilde{m})}{\Gamma (1+m)}\int \frac{%
d^{2}y\,\,K(\overrightarrow{x},\overrightarrow{y})}{\pi \,\left|
y(1-y)\right| ^{2}}\,\,\left( \varphi
_{m,\widetilde{m}}(\overrightarrow{y}%
)\right) ^{*}\,,
\end{equation}
where for even conformal spins $n=m-\widetilde{m}$ the Green function $K(%
\overrightarrow{x},\overrightarrow{y})$ contains the contribution of zero
modes $t$: 
\begin{equation}
K(\overrightarrow{x},\overrightarrow{y})=(x-y)^{m}(x^{*}-y^{*})^{\widetilde{
m%
}}-\frac{t(\overrightarrow{x})\,t(\overrightarrow{y})}{6}\,\,,\,\,t(%
\overrightarrow{x})=x^{m}x^{*\widetilde{m}}+(1-x)^{m}(1-x^{*})^{\widetilde{m
}%
}+1\,.
\end{equation}

If we use the definitions 
\begin{equation}
a_{m}=\,X\,p^{m+1},\,\,\,\,a_{1-m}=\,X\,\,p^{2-m}\,,\,\,\,X\equiv
x(1-x)\,\,
\end{equation}
the third order differential equation $q _3 f_{m}=\lambda _{m}f_{m}$ [1] for
holomorphic factors is equivalent to the system of two pseudo-differential
equations 
\begin{equation}
a_{m}\,\varphi _{m}=l_{m}\,\,\varphi _{1-m}\,\,,\,\,\,a_{1-m}\,\varphi
_{1-m}=l_{1-m}\,\varphi _{m}\,,\,\,l_{1-m}l_{m}=\lambda _{m}\,.
\end{equation}

The function $\varphi _{1-m}(x)$ has the conformal weight equal to $1-m$.
It
is important, that there is another relation between $\varphi _{m}$ and $%
\varphi _{1-m}$%
\begin{equation}
\varphi _{1-m}=R_{m}(x,p)\,\varphi
_{m}\,,\,R_{m}(x,p)=X^{-m+1}\,p^{1-2m}\,X^{-m}\,,\,\,R_{1-m}(x,p)=\left(
R_{m}(x,p)\right) ^{-1}\,.
\end{equation}
It follows from the fact, that for the M\"{o}bius group the complex
conjugated representations $O_{m,\widetilde{m}}(\overrightarrow{\rho
_{0}})$
and $O_{1-m,1-\widetilde{m}}(\overrightarrow{\rho _{0}})$ are linearly
dependent.

There are three independent solutions $\varphi _{i}^{(m)}(x,\lambda )$ of
the third order differential equation $a_{1-m}\,a_{m}\,\varphi =\lambda
\varphi$ \,for each eigenvalue $\lambda $. If we present them as the
power series in $x$, starting with the terms: 
\begin{equation}
\varphi _{r}^{(m)}(x,\lambda )=x+O(x^{2})\,,\,\,\varphi
_{s}^{(m)}(x,\lambda
)=\frac{i}{\lambda }(m-1)+x\,\ln x\,+\,O(x^{2})\,,\,\,\varphi
_{f}^{(m)}=x^{m}+O(x^{m+1}),
\end{equation}
the equation is reduced to simple recurrence relations for the
coefficients.
The appearance of $\ln x$ in $\varphi _{s}^{(m)}(x,\lambda )$ is related
with the degeneracy of the differential equation.

{}From the point of view of the operator product expansion for the gluon
fields
at $\rho _{1}\!\rightarrow \!\rho _{2}$ one can consider the functions $\varphi
_{i}^{(m)}(x,\lambda )$ as contributions of the series of holomorphic 
composite operators starting with the ones having the conformal weights
$M=0,\,m$
and $1$ for $i=s,\,f$ and $r$ correspondingly. In this interpretation the
above degeneracy is related with the existence of the conserved vector
current (for $m=1/2$ there is also a conserved fermion current).

The single-valuedness condition for the Odderon wavefunction at $%
\overrightarrow{x}=0$ and the modular invariance under the transformation
$x
\rightarrow -x/(1-x)$ lead to its holomorphic factorization representation
in the form

\[
\varphi _{m,\widetilde{m}}(\overrightarrow{x})=\varphi _{f}^{(m)}(x,\lambda
)\,\varphi _{f}^{(\widetilde{m})}(x^{*},\lambda ^{*})+\varphi
_{f}^{(m)}(x,-\lambda )\,\varphi _{f}^{(\widetilde{m})}(x^{*},-\lambda
^{*}) 
\]
\[
+c_{1}\left( \varphi _{s}^{(m)}(x,\lambda )\,\varphi _{r}^{(\widetilde{m}%
)}(x^{*},\lambda ^{*})+\varphi _{r}^{(m)}(x,\lambda )\,\varphi _{s}^{(%
\widetilde{m})}(x^{*},\lambda ^{*})+\left( \lambda \rightarrow -\lambda
\right) \right) 
\]
\begin{equation}
+c_{2}\left( \varphi _{r}^{(m)}(x,\lambda )\,\varphi _{r}^{(\widetilde{m}%
)}(x^{*},\lambda ^{*})+\varphi _{r}^{(m)}(x,-\lambda )\,\varphi _{s}^{(%
\widetilde{m})}(x^{*},-\lambda ^{*})\right) .
\end{equation}

The complex coefficients $c_{1},c_{2}$ and eigenvalues $\lambda $ are
fixed 
{}from the single-valuedness conditions for $\varphi _{m,\widetilde{m}}(%
\overrightarrow{x})$ at the points $\overrightarrow{x}=1$ and $%
\overrightarrow{x}=\infty $ [4,5]. From the duality equation and the reality
condition for the M\"{o}bius group representation we have the following
relations among these coefficients: 
\begin{equation}
\left| c_{1}\right| =\left| \lambda \right| \,,\,\,\mbox{Im} \,\frac{c_{2}}{%
c_{1}}=
\mbox{Im}
\,(m^{-1}+\widetilde{m}^{-1})=-4\nu 
\left[\frac{1}{4\nu ^2+(n+1)^2}+\frac{1}{4\nu ^2+(n-1)^2}\right]\,. 
\end{equation}
One can verify from the numerical results of papers [5], that these
relations are fulfilled. After the Fourier transformation of $\varphi _{m,%
\widetilde{m}}(\overrightarrow{x})$ to the momentum space
$\overrightarrow{p}
$ the regular terms near points $\overrightarrow{x}=0$ and
$\overrightarrow{x%
}=1$ do not give any contribution to the asymptotic behaviour of $\varphi$ 
at $\overrightarrow{p}\rightarrow \infty $ and therefore $\varphi
_{m,\widetilde{%
m}}(\overrightarrow{p})$ near $\overrightarrow{p}=\infty $ can be
calculated
as an asymptotic series without any additional parameters except $\lambda
$. In particular for $m=\widetilde{m}=1/2$ the Odderon wavefunction 
satisfies a non-relativistic Schr\"{o}dinger equation with the potential 
$V(p)=p^{-3/2}$.
The requirements of the holomorphic factorization and single-valuedness in 
the momentum space lead
to the quantization of $\lambda $.

{}From the duality and reality conditions one can derive also integral
representations for the coefficients $c_{1,2}$ in terms of integrals of
$%
\varphi _{m,\widetilde{m}}(x,x^{*})$ over the fundamental region of the
modular group, where the expansion in $x$ is convergent. These relations
allow to calculate the coefficients $c_{1,2}$ without  the
single-valuedness condition.

Using the relation of $q_{3}$ with pair Kasimir operators of the
M\"{o}bius group 
\begin{equation}
q_ 3 =\frac{i^{3}}{4}\left[ \overrightarrow{M}^{2}\,,\,\left(
\overrightarrow{M}%
_{3},\overrightarrow{N}\right) \right] \,,\,\,\overrightarrow{M}=%
\overrightarrow{M}_{1}+\overrightarrow{M}_{2}\,\,,\,\,\overrightarrow{N}=%
\overrightarrow{M}_{1}-\overrightarrow{M}_{2}\,,\,\,
\end{equation}
its matrix elements can be expressed in the representation where $%
\overrightarrow{M}^{2}$ is diagonal through the matrix elements of the
operator $\overrightarrow{N}$ which are known because $\overrightarrow{M}$
and $\overrightarrow{N}$ are generators of the Lorentz group.

The functions $\varphi _{M}^{m}(x)$
having the fixed total gluon momentum $m$ and the fixed pair momentum $M$ 
can be constructed in terms of hypergeometric functions

\begin{equation}
\varphi _{M}^{m}(x)=x^{M}(1-x)^{m}\,F(m+M,M,2M;x).
\end{equation}

Using the fact, that they satisfy the following relations: 
\begin{equation}
q _3 \,\varphi _{M}^{m}\,(x)=-\frac{i^{3}}{2}M(M-1)(M-m)\left( 
\frac{(M+m)(M+m-1)}{%
2(2M-1)(2M+1)}\,\varphi _{M+1}^{m}(x)-2\varphi _{M-1}^{m}(x)\right) ,
\end{equation}
we can present the three independent solutions of the equation
$q _3 \,\,\varphi
(x)=\lambda \,\varphi (x)$ as the superposition of $\varphi _{M}^{m}(x)$ 
with the
coefficients satisfying recurrence relations. This representation is
better 
than the above power series in $x$, because it gives a possibility to
calculate also the eigenvalue of $h$ as a function of $\lambda $ in a way 
different from one suggested in ref. [6].

In conclusion we note that the remarkable properties of the reggeon
dynamics are related presumably to the extended supersymmetry. An 
argument supporting this point of view was given in [7]. Namely, the
eigenvalues of the integral kernels in the evolution equations for
quasi-partonic operators in the $N=4$ SUSY are proportional to $\psi (j-1)$,
which means that these evolution equations in the multi-colour limit are
equivalent to the Schr\"odinger equation for the integrable Heisenberg spin
model similar to that found in the Regge \mbox{limit [8].}

\vspace{.4cm}
\noindent
{\bf Acknowledgements}\\
I would like to thank L.\ Faddeev, E.\ Antonov, A.\ Bukhvostov, P.\ Gauron, 
B.\ Nicolescu, J.\ Bartels, R.\ Kirschner, L.\ Szymanowski, 
R.\ Janik, J.\ Wosiek, 
M.\ Braun and  C.\ Ewerz for helpful discussions.

\end{document}